\documentclass[aip,jcp,amsmath,amssymb,reprint,numeric]{revtex4-1}
\usepackage{graphicx}
\usepackage{subfigure}
\usepackage{color}
\usepackage{amsmath}
\usepackage{bm}

\begin{document}
\title {Composition dependence of the glass forming ability in binary mixtures: The role of demixing entropy}
\author{Ujjwal Kumar Nandi}
\affiliation{\textit{Polymer Science and Engineering Division, CSIR-National Chemical Laboratory, Pune-411008, India}}
\author{Atreyee Banerjee}

\affiliation{\textit{Polymer Science and Engineering Division, CSIR-National Chemical Laboratory, Pune-411008, India}}

\author{Suman Chakrabarty}
\affiliation{\textit{Physical and Materials Chemistry Division, CSIR-National Chemical Laboratory, Pune-411008, India}}

\author{Sarika Maitra Bhattacharyya}
\email{mb.sarika@ncl.res.in}
\affiliation{\textit{Polymer Science and Engineering Division, CSIR-National Chemical Laboratory, Pune-411008, India}}

\date{\today}

\date{\today}

\begin{abstract}
We present a comparative study of the glass forming ability of binary systems with varying composition, where the systems have similar global crystalline structure (CsCl+fcc). Biased Monte Carlo simulations using umbrella sampling technique shows that the free energy cost to create a CsCl nucleus increases as the composition of the smaller particles are decreased. We find that the systems with comparatively lower free energy cost to form CsCl nucleus exhibit more pronounced pre-crystalline demixing near the liquid/crystal interface. The structural frustration between the CsCl and fcc crystal demands this demixing. We show that closer to the equimolar mixture the entropic penalty for demixing is lower and a glass forming system may crystallize spontaneously when seeded with a nucleus. This entropic penalty as a function of composition shows 
a non-monotonic behavior with a maximum at a composition similar to the well known Kob-Anderson (KA) model. Although the KA model shows the maximum entropic penalty and thus maximum frustration against CsCl formation, it also shows a strong tendency towards crystallization into fcc lattice of the larger ``A" particles which can be explained from the study of the energetics. Thus for systems closer to the equimolar mixture although it is the requirement of demixing which provides their stability against crystallization, for KA model it is not demixing but slow dynamics and structural frustration caused by the locally favored structure around the smaller ``B" particles which make it a good glass former. Although the glass forming binary systems studied here are quite similar, differing only in composition, we find that their glass forming ability cannot be attributed to a single phenomena.
 
\end{abstract}
\maketitle
\section{Introduction}
A liquid upon cooling undergoes first order phase transition and forms a crystal. However if the cooling rate is increased
 it cannot crystallize and forms an amorphous glassy material \cite{kob_binder}. In addition to fast supercooling, there are other methods to favor glass
 formation over crystallization. In bulk metallic glass community the usual thumb rules are to at least have a two component mixture with negative enthalpy
 of mixing and a 12$\%$ size ratio between the components \cite{Acta_mater}.  Single component systems are known to crystallize in a fcc+hcp structure
 \cite{Hoover_Ree}, thus multi-component systems are commonly used for making glasses. The negative enthalpy of mixing makes sure that the components
 remain in a mixed state and do not demix to form single component crystals, whereas the size ratio provides frustration in packing. Although there is an array
 of experimental systems which form glasses, in computer simulation studies there is only a handful of systems known to be good glass former
 \cite{kob,wahnstrom,NTW,ohern,harowell_cuzr_nature_let}. Note that most of the glass forming systems have global crystalline minima  \cite{harrowell-jcp, paddyal2cu}. Thus depending on the barrier to crystallization it is just a matter of time for the systems to crystallize. With the increase in the available computational power some of the well known glass former like Kob-Anderson (KA) model and Wahnstrom (WA) model are now found to crystallize
 \cite{dyre,pedersen}. Thus in order to design better glass formers we need to be able to estimate the glass forming ability (GFA) 
of these systems.

In order to quantify GFA, first we need to understand the origin behind the stability against crystallization. This is an active
 field of research and different studies have attributed the GFA to different phenomena \cite{frank,tarjus,tarjus_PRL_curvedspace,tanaka-jnoncryst2,tanaka-nature12,paddyal2cu,tanaka-jnoncryst1}. The most popular among them is the theory of frustration first proposed by Frank \cite{frank}. According to him, the local liquid ordering is different from the
 crystalline order and this frustrates
 the system and decreases the rate of crystallization.
It has also been argued that regions with locally favored structures (LFS) give rise to domains and are connected to the slow dynamics in the supercooled liquids 
\cite{tarjus,tarjus_PRL_curvedspace}. Sometimes the LFS can also be related to the underlying crystalline structure
 \cite{tanaka-jnoncryst2,tanaka-nature12,paddyal2cu}. In some cases the LFS connected to crystal structure grows more than the one connected to the
 liquid structure \cite{tanaka-nature12}. The locally favored structures can be different in different dimensions. There are LFS, like the icosahedral
 ordering, which can cause frustration in the Euclidean space but tile the curved space \cite{tarjus_PRL_curvedspace}. Frustrations are not always structural
 but can also be energetic in nature \cite{tanaka-jnoncryst1}.

Most binary equimolar mixtures form crystalline structures \cite {vlot}, where
the crystal structure may vary according to the size ratio of the components. There are also some exceptions like the equimolar CuZr structure which is found 
to be a good glass former \cite{harowell_cuzr_nature_let}. However, when the composition of the mixtures are changed then it is usually found that close to
the deep eutectic point many of them form glasses. One of the argument in favor of the deep eutectic point being a good glass forming zone is that 
the viscosity is highest at this point so kinetically it takes a longer time to form a crystal nucleus.  However it has also been shown that the structural 
frustration between two different crystal structures can make this region a good glass former.
 This kind 
of phase diagram (in temperature vs. composition space) 
are often referred to as a V-shaped phase diagram where the bottom of the V is the glass forming region \cite{tanaka-epje,tanaka-vshaped-prl,valeria,charu_vshaped,atreyee}.

In a recent work by some of us we have shown that even though all the systems at equimolar mixture undergo 
crystallization, as the composition of the larger size particles increases, the zone which forms CsCl crystal at equimolar composition does not crystallize any more
 \cite{atreyee}. It is already known from the study of energetics that the global free energy minima of these systems are CsCl+fcc crystals \cite{harrowell-jcp}. 
The well known KA glass former is one of the systems present in this more generic CsCl zone and it has shown strong resistance towards crystallization even after being inserted with a CsCl seed in the liquid \cite{harowell}. So far only in one study it has been reported to  crystallize but in a structure which is different from that of
the global minima \cite{dyre}. 
In the earlier study we have shown that in the CsCl+fcc crystal structure the bigger ``A" particles need to have two different populations where there is a large difference in the order parameter (coordination number and bond orientational order parameter) of these two populations. According to us, this large difference in order parameter creates frustration. Thus the stability against crystallization is attributed to the structural frustration between the CsCl and the fcc crystal structure \cite{atreyee}.    

In this present work we study a similar series of binary systems by changing the composition and also the inter species interaction length. Many of the binary systems studied here are good glass formers and have a global minima which is CsCl+fcc structure. Thus according to our earlier study the structural frustration for these systems are similar. However, these different systems although share the same structural frustration are expected to have different glass forming ability. The goal of this work is to get a relative estimate of the GFA of different systems and then explore the origin behind their differences. Our study shows that the free energy cost for CsCl crystallization increases with the composition of the smaller particles. The system with lowest free energy cost also shows a pre-crystalline demixing in the liquid phase near the liquid/crystal interface. The demixing takes place due to the structural frustration between the CsCl and fcc structures. Upto a certain composition, the composition dependence of the free energy cost to create a crystal nucleus can be related to the composition dependence of this demixing entropy. Our study of energetics shows that although in the whole range of composition the global 
minima is CsCl+fcc crystal the driving force of crystallization in a certain region is the CsCl crystal and in another region is fcc crystal. In the former region the system tends to demix and form CsCl+fcc crystal and demixing frustrates the crystallization process. However, in the latter region we show that demixing does not play a crucial role. It is primarily the slow dynamics near eutectic point and LFS around the smaller ``B" particles which frustrate the crystallization process.

The simulation details are given in the next section. In section III we present the definition and method for evaluating different quantities, in section IV we have the results and discussion, and section V ends with a brief summary.

\section{Simulation Details}
The atomistic models which are simulated are two component mixtures of classical
particles (larger ``A'' and smaller ``B'' type), where particles of type 
{\it i} interact with those of type {\it j} with pair 
potential, $U_{ij}(r)$,  where r is the distance between the pair. 
$U_{ij}(r)$ is described by a shifted and truncated Lennard-Jones (LJ) potential, as given by:
\begin{equation}
 U_{ij}(r)=
\begin{cases}
 U_{ij}^{(LJ)}(r;\sigma_{ij},\epsilon_{ij})- U_{ij}^{(LJ)}(r^{(c)}_{ij};\sigma_{ij},\epsilon_{ij}),    & r\leq r^{(c)}_{ij}\\
   0,                                                                                       & r> r^{(c)}_{ij}
\end{cases}
\end{equation}

\noindent where $U_{ij}^{(LJ)}(r;\sigma_{ij},\epsilon_{ij})=4\epsilon_{ij}[({\sigma_{ij}}/{r})^{12}-({\sigma_{ij}}/{r})^{6}]$ and
 $r^{(c)}_{ij}=2.5\sigma_{ij}$. Subsequently, we'll denote A and B types of particles by indices 1 and 2, respectively.

The different models are distinguished by different choices of lengths and composition parameters. Length, temperature and
time are given in units of $\sigma_{11}$, ${k_{B}T}/{\epsilon_{11}}$ and $\surd({m_{1}\sigma_{11}^2}/{\epsilon_{11}})$, 
respectively.  
Here we have simulated various binary mixtures  
with the interaction parameters  $\sigma_{11}$ = 1.0, $\sigma_{22}$ =0.88 , $\epsilon_{11}$ =1, $\epsilon_{12}$ =1.5,
 $\epsilon_{22}$ =0.5, $m_{1} =m_{2}=1$ and the inter-species interaction length $\sigma_{12}=0.7,0.8$. 
 We have 
simulated systems with different compositions, varying $x_B$ from 0.50 to 0.0, where $x_B$ is the mole fraction of the smaller B type particles \cite{dyre,valdes-jcp}. 

The molecular dynamics (MD) simulations have been carried out using the LAMMPS 
package \cite{lammps}.
We have performed MD simulations in the isothermal−isobaric ensemble (NPT) using  Nos\'{e}-Hoover thermostat and Nos\'{e}-Hoover barostat with integration timestep 0.005$\tau$. The time
constants for  Nos\'{e}-Hoover thermostat and barostat are taken to be 100 and 1000 timesteps, respectively.
Except for the liquid/crystal interface study where we use a rectangular box, all of the other studies are performed in a cubic box with periodic boundary condition. The free energy barrier calculations are done via biased Monte Carlo method. All the studies are performed at $P=0.5$.

\section{Definitions}
\subsection {Bond Orientational Order Parameter}
Bond  Orientational Order (BOO) parameter was 
first prescribed by
Steinhardt {\it et al.}  to characterize specific crystalline structures \cite{steinhardt}. 
To characterize specific crystal structures  we have calculated the locally averaged BOO 
parameters  ($\bar{q}_{lm}$) of \textit{l}-fold symmetry as a 2\textit{l}+1 vector,\cite{dellago} 
\begin{eqnarray}
  \bar{q}_l =\sqrt{\frac{4\pi}{2l+1}\sum_{m=-l}^{l}\arrowvert \bar{q}_{lm}\arrowvert^2} 
\end{eqnarray}
\noindent
where 
\begin{eqnarray}
\bar{q}_{lm}(i)=\frac{1}{\tilde{N_i}} \sum_{0}^{\tilde{N_{i}}} q_{lm}(k).
\end{eqnarray}
\noindent
Here $\tilde{N_{i}}$ is the number of neighbours of the i-th particle and the particle i itself. $q_{lm}(i)$ is the local BOO of the i-th particle. 
\begin{eqnarray}
 q_{lm}(i)=\frac{1}{N_i} \sum_{0}^{N_{i}} Y_{lm}(\theta(r_{ij}), \phi(r_{ij})) 
\end{eqnarray}
where $Y_{lm}$ are the spherical harmonics,  $\theta(r_{ij})$ and $\phi(r_{ij})$ are 
spherical coordinates of a bond $r_{ij}$ in a fixed reference frame,
and $N_i$ is the number of neighbours of the {\it i}-th particles. Two particles are considered neighbours if $r_{ij}< r_{min}$, where $r_{min}$ is the first minimum 
of the radial distribution
function (RDF). For the liquids and the crystals the $r_{min}$ has been chosen as the first minima of the respective partial RDF of the ``A" type of particles.  For the pure CsCl crystal this comprises of 14 neighbours and for fcc 12 neighbours. 

\begin{figure}[h]
\centering
\includegraphics[width=.45\textwidth]{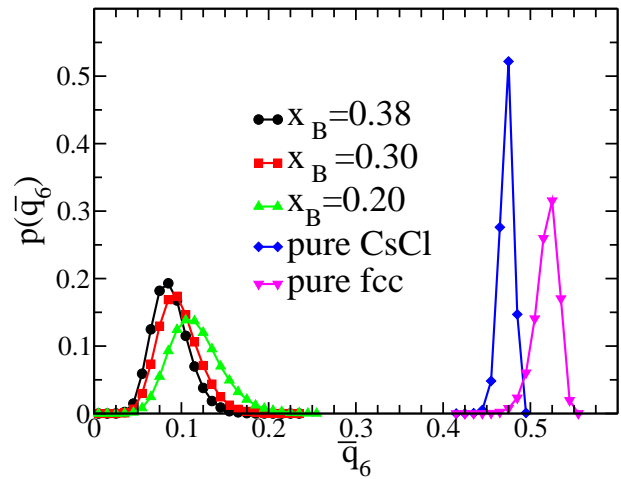}
\caption{ The probability distribution of the locally averaged BOO, $\bar{q}_{6}$ for the liquid at three different compositions $x_{B}=0.38, 0.3, 0.2$ at $T=0.5$. We also plot 
the same for the CsCl crystal made up of ``A" and ``B" type of particles and pure fcc made up of ``A" particles.} 
\label{fig1}
\end{figure}

In Fig.\ref{fig1} we plot the probability distribution of $\bar{q}_{6}$ of the liquid at three different composition and also the same for pure CsCl and fcc crystals. 
We note that at the level of this parameter all the three liquids can be clearly separated from the two different crystal forms. 


\subsection{Relaxation Time}
We have calculated the relaxation times obtained from the decay of the
overlap function $q(t)$, where $q(t = \tau_{\alpha} , T )/N =
1/e$. It is defined as
\begin{eqnarray}
\langle q(t) \rangle \equiv \left \langle \int dr \rho(r, t_0 )\rho(r, t + t_0 )\right \rangle \nonumber\\
=\left \langle \sum_{i=1}^{N}\sum_{j=1}^{N} \delta({\bf{r}}_j(t_0)-{\bf{r}}_i(t+t_0)) \right \rangle \nonumber\\
=\left \langle \sum_{i=1}^{N} \delta({\bf{r}}_i(t_0)-{\bf{r}}_i(t+t_0)) \right \rangle \nonumber\\
+\left \langle \sum_{i}\sum_{j\neq i} \delta({\bf{r}}_i(t_0)-{\bf{r}}_j(t+t_0)) \right \rangle
\end{eqnarray}
The overlap function is a two-point time correlation
function of local density $\rho(r, t)$. It has been used in
many recent studies of slow relaxation \cite{unravel}.
 In this
work, we consider only the self-part of the total overlap
function (i.e. neglecting the $i \neq j$ terms in the double
summation). Earlier it has been shown to be a good approximation to the full overlap function. So, the self part of the overlap function can be written as,
\begin{eqnarray}
 \langle q(t) \rangle \approx \left \langle \sum_{i=1}^{N} \delta({\bf{r}}_i(t_0)-{\bf{r}}_i(t+t_0)) \right \rangle
\end{eqnarray}

Again, the $\delta$ function is approximated by a window function $\omega(x)$ which defines the
condition of “overlap” between two particle positions
separated by a time interval t:
\begin{eqnarray}
 \langle q(t) \rangle \approx \left \langle \sum_{i=1}^{N} \omega (\mid{\bf{r}}_i(t_0)-{\bf{r}}_i(t+t_0)\mid) \right \rangle \nonumber\\
\omega(x) = 1, x \leq a {\text{implying ``overlap''}} \nonumber\\
=0, \text{otherwise}
\end{eqnarray}

The time dependent overlap function thus depends on
the choice of the cut-off parameter $a$, which we choose
to be 0.3. This parameter is chosen such that particle positions separated due to small amplitude vibrational motion are treated as the same, or that $a^2$ is comparable to
the value of the MSD in the plateau between the ballistic
and diffusive regimes.

\section{Results}
 
\subsection{Melting Temperatures}
\begin{figure}[h]
\centering
\includegraphics[width=.45\textwidth]{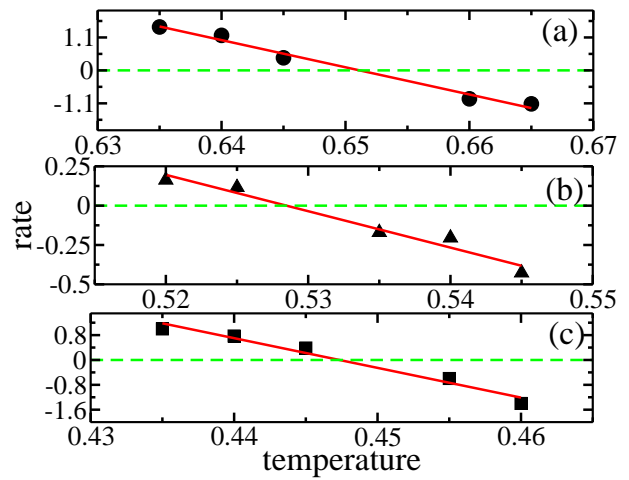}
\caption{The growth rate (negative for melting) of the pure CsCl crystal as a function of temperature for different compositions: (a) For the $x_{B}=0.38$ system, the predicted melting temperature ($T_m$) is 0.651. (b) $x_{B}=0.3$, $T_m = 0.528$. (c) $x_{B}=0.2$, $T_m =  0.447$.} 
\label{fig2}
\end{figure}

\begin{figure}[h]
\centering
\includegraphics[width=.45\textwidth]{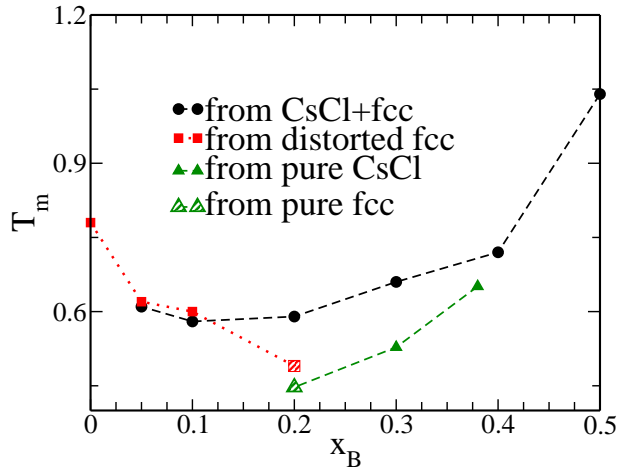}
\caption{ The melting temperatures of different crystal forms in different binary mixtures. The melting temperature for the pure fcc and CsCl crystal
 is obtained by studying the growth/melt rate of the crystal. The melting temperature is where the predicted rate disappears (see Fig \ref{fig2}). The melting 
temperature of the mixed CsCl+fcc crystal and the distorted fcc crystal are obtained by step wise heating the system. } 
\label{fig3}
\end{figure}

In order to calculate the crystallization rate and thus the glass forming ability we first determine the melting temperatures of the different crystals.
The melting temperature is studied by calculating the temperature dependent growth/melting rate of the crystal and  fitting them to a straight line. The temperature at which the growth rate cuts the temperature axis is the predicted melting temperature where the growth rate goes to zero \cite{paddyal2cu}.  The simulations are done
 at $P=0.5$. With the crystal at the center of the box and the crystal particles being pinned the liquid of 8000 particles is equilibrated at T=1.5. The system is then quenched to the target lower
 temperatures and the crystal particles are unpinned. We then run a short equilibration of 1000 steps for the quenched system. Depending on the temperature and the
 composition of the liquid the central seed either grows or melts. In the $x_{B}=0.38$ and $0.3$ systems we study the melting temperature of CsCl crystal with an initial
 crystal seed of 432 particles. In the $x_{B}=0.2$ mixture we study the melting temperature of the pure fcc crystal comprising of 500 ``A" particles. 
The growth of the seed is monitored by cluster analysis where the $\bar{q}_{6}$ is calculated for each particle and if the value of $ \bar {q}_{6} >0.3$ (Fig.\ref{fig1}) 
and it has a neighbour which is part of the existing cluster then it is included in the cluster. The cluster growth is monitored for about 100-500 $\tau_{\alpha}$,
 where $\tau_{\alpha}$ is the temperature dependent $\alpha$ relaxation time that varies across different systems. 5-10 independent runs are generated at each temperature 
by starting from the same initial configuration but randomized initial velocity. The growth rate is calculated by scaling the time w.r.t the corresponding $\tau_{\alpha}$. From the average growth/decay rate we approximate the melting temperature as the
 temperature  where the predicted growth or decay rate is zero (Fig.\ref{fig2}).  The melting points obtained from Fig.\ref{fig2} is used to construct the composition dependent phase diagram reported in Fig.\ref{fig3}.
  
\begin{figure}[h]
\centering
\subfigure{
\includegraphics[width=0.3\textwidth]{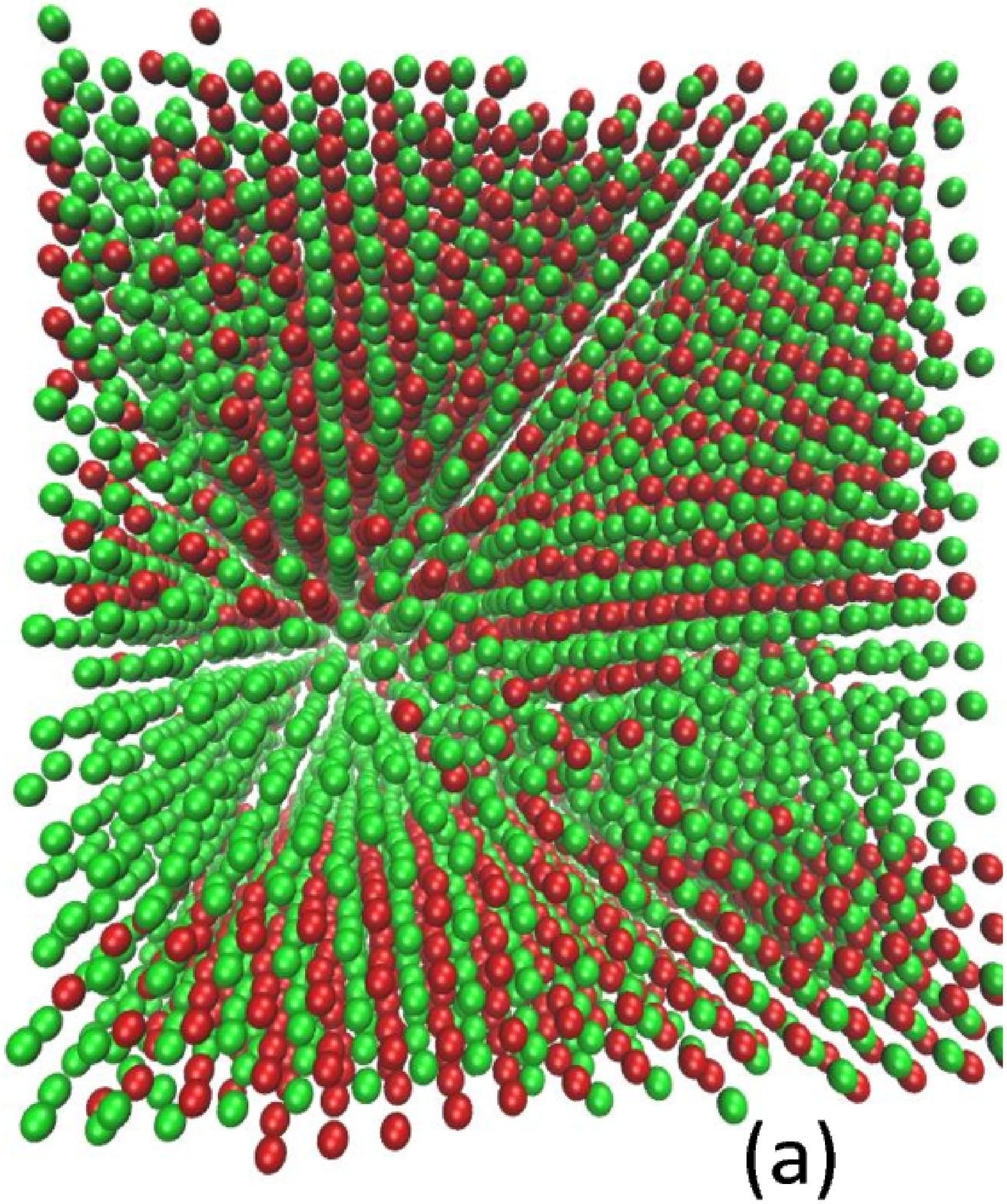}}
\subfigure{
\includegraphics[width=0.3\textwidth]{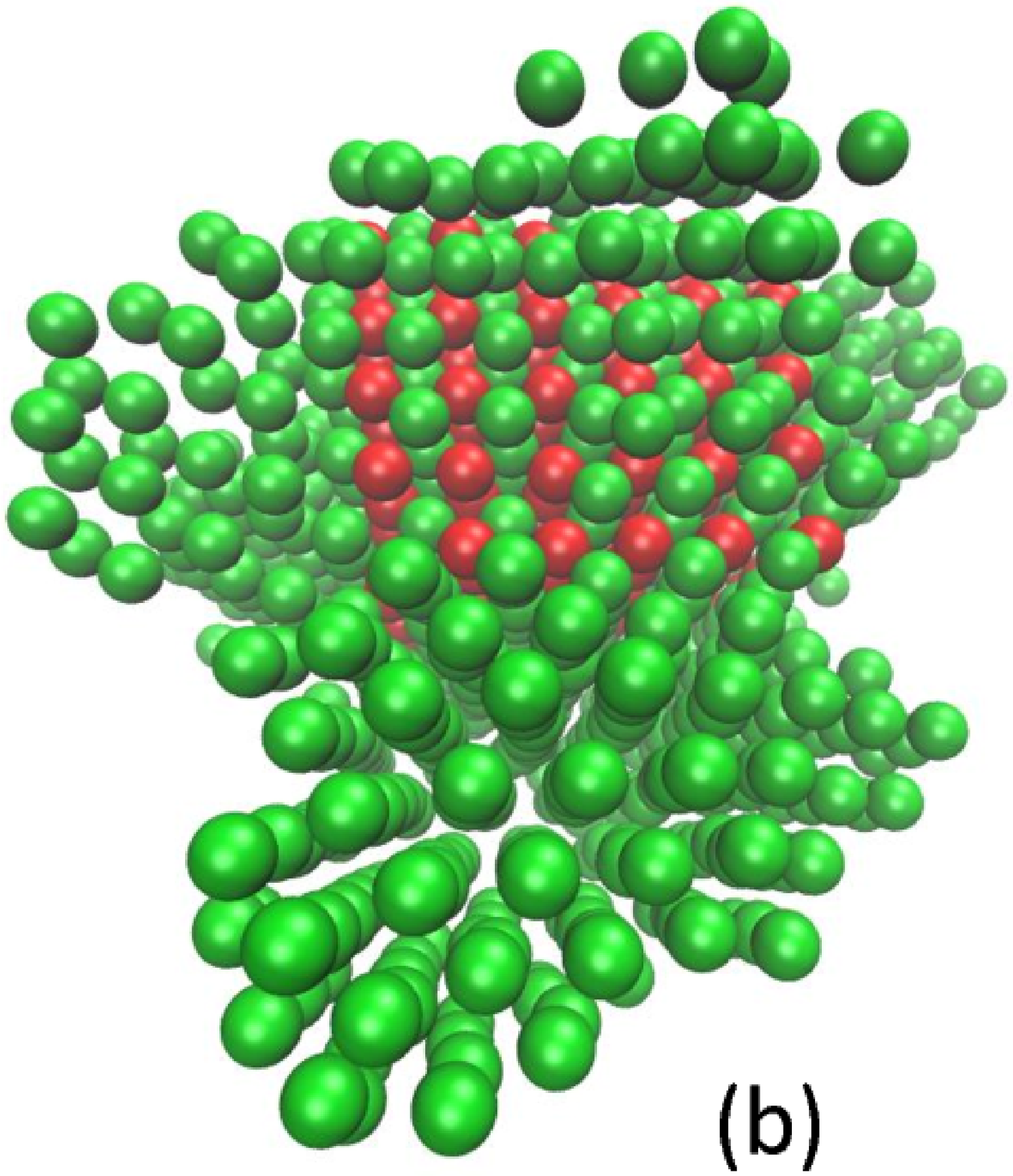}}
\caption{(a) Snapshot of the $x_{B}=0.38$ system after $10^{7}$ steps post quench at temperature 0.52. The system shows clear demixing and grows into CsCl+fcc crystal. 
(b) Snapshot of the initial seed and the cluster that grew around it for $x_{B}=0.2$ system after $2\times10^{7}$ steps post quench at temperature 0.4. Around the CsCl seed we find the growth of fcc crystal of ``A" particles. For both the systems the initial CsCl crystal seed consists of 432 particles which is inserted in a liquid of 8000 particles.}
\label{fig4}
\end{figure}

We find that the $x_{B}=0.38$ mixture phase separates and forms a CsCl+fcc crystal structure (Figs.\ref{fig4}a). The $x_{B}=0.3$ mixture also
 shows similar tendency however the crystal growth rate is slower and within our simulation timescale the demixing is not complete.
We also try to grow the CsCl crystal in the $x_{B}=0.2$ mixture but we find that instead of CsCl , fcc structure of ``A" particles grow around the initial seed 
(Fig.\ref{fig4}b). This is similar to the observation reported earlier \cite{harowell}. When a fcc seed is inserted in the same mixture it continues to grow.

In the above mentioned method it is not possible to calculate the melting temperature of the mixed CsCl+fcc crystal as the growth of such crystal never happens. 
 For this calculation at each composition ($x_{B}=0.05-0.5$) we heat the mixed crystal (CsCl+fcc) starting from temperature 0.2-0.3 and increase it up to 0.59-1.0 
(depending on the melting temperature of the crystal) with temperature interval of 0.05. Closer to the melting temperature heating is done with 0.01 temperature 
interval. At each temperature equilibration is done for $10^7$ steps. The size of the initial crystal structure is in the range of 468-612.  The total number of
 particles are chosen in such a way that a perfect mixed crystal can be created. Periodic boundary condition is applied in all directions.
 Similar study is been done for the pure and distorted fcc crystal for $x_{B}=0.0,0.1,0.2$ systems.
For $x_{B}=0.0$ we get pure fcc and for $x_{B}=0.05$ and $0.01$ the ``A" particles form fcc crystal but with distortion due to presence of the 
``B" particles. In the $x_{B}=0.2$ system within our simulation run we could not form the fcc crystal. However as reported earlier in a MKA2 model, if the interaction between
 the two species is reduced, then the system forms crystal \cite{dyre}. In a similar method by keeping the $\epsilon_{12}=0.96$ we first form a distorted fcc crystal
 of the $x_{B}=0.2$ system. Once the crystal is formed we switch back to the larger inter species interaction of $\epsilon_{12}=1.5$ and study its melting. 
 The melting of all the crystals happen instantaneously. The melting temperatures are reported in Fig.\ref{fig3}. 

\subsection{Free Energy of Nucleation and Role of Demixing}

\begin{figure}[h]
\centering
\includegraphics[width=.45\textwidth]{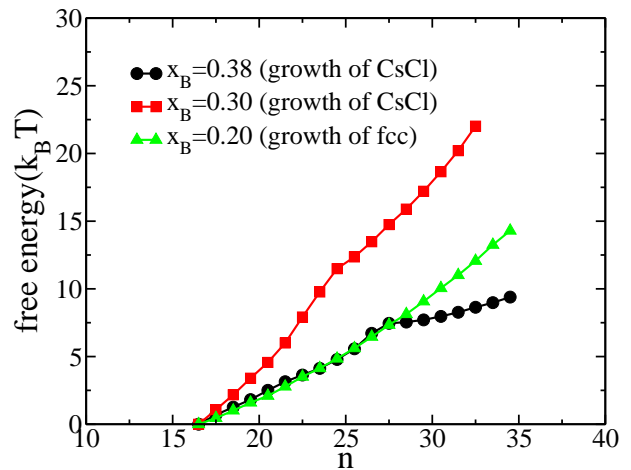}
\caption{Gibbs free energy for crystallization as obtained from the umbrella sampling simulations as a function of the largest cluster
 size for the three systems. For $x_{B}=0.38$ and $0.3$ we can grow the CsCl cluster, whereas for $x_{B}=0.2$ we can only grow the fcc cluster. Even with a initial 
small CsCl seed the cluster that grows is of ``A" particles forming fcc lattice which is similar to that we find for melting study.} 
\label{fig5}
\end{figure}

In this section we perform a comparative study of the Gibbs free energy (potential of mean force) of crystalline nucleation/growth in different binary mixtures using umbrella sampling technique with the reaction coordinate being the size of the largest crystalline cluster present in the system.
The studies are performed at the same degree of undercooling at $0.8T_{m}$, where the melting temperatures used are those calculated by studying the 
temperature dependent growth/melting rate for the pure CsCl and fcc crystals. 
A crystalline cluster is 
defined by a neighborhood criteria (within a cut-off distance determined by the first minimum of the partial radial distribution of function of ``A''-type particles for respective systems) of ``crystal-like'' particles (with the criterion of $\bar{q}_{6} > 0.3$). To grow the clusters we use a biased Monte Carlo approach, where we apply
 an external harmonic potential of the form $\frac{1}{2}k(n-n_{c})^{2}$, where $k$ is the force constant, $n$ is the number of particles in the largest cluster, and $n_{c}$ is the position of the bias window. We use $k=0.1$ for $x_{B}=0.38$, and $k=0.2$ for $x_{B}=0.30$ and $x_{B}=0.20$. We have used 5-7 umbrella windows (depending on the system) in the cluster size range of 15-35. After equilibration, the data is collected for
$10^4$ Monte Carlo steps per window and Weighted Histogram Analysis method (WHAM)\cite{wham}  is then used to
 compute the free energy as a function of the size of the largest cluster as reported in Fig. \ref{fig5}.
 
\begin{figure}[h]
\centering
\includegraphics[width=.45\textwidth]{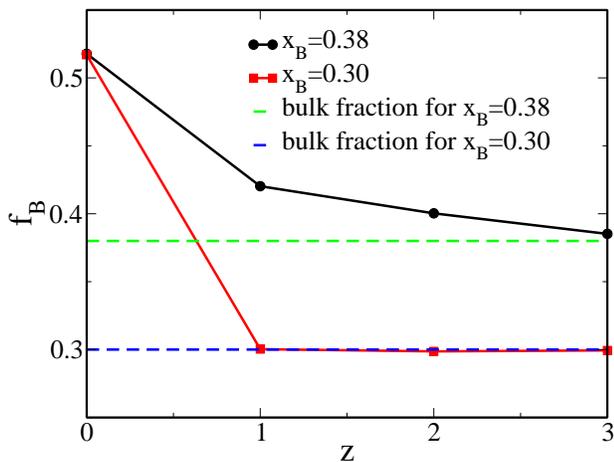}
\caption{Distance dependence of the composition from the liquid/crystal interface along the perpendicular direction. We plot the fraction of ``B" particles, $f_{B}$, 
as obtained within each slab of width one $\sigma_{11}$ as a function of the distance from the interface. The interface that has ``A" particles is taken 
and the plot is done for $x_{B}=0.38$ and $x_{B}=0.3$. We find that for the former system where the initial rate of crystallization is higher the interface
 has higher concentration of ``B" particles compared to the bulk. Thus there is pre-crystalline demixing in the liquid phase.} 
\label{fig6}
\end{figure}

While our calculations focus on the pre-critical region of the free energy surfaces, we can compare the relative free energy cost to form a crystalline nucleus of certain size as the composition of the system is varied.
We observe that the free energy cost to grow a nucleus from 15 to 35 for all the systems are quite high (in the range of 10-20 $k_{B}T$), which explains why all these systems are good glass formers.
 A comparative study of the cost of free energy shows that $x_{B}=0.38$ has a lower cost to grow a CsCl crystal compared to $x_{B}=0.3$. This explains 
the slow growth of the CsCl crystal in the latter system which is observed during the melting study. 
We also try to grow CsCl crystal for $x_{B}=0.2$, which we do not observe during our simulation time. This implies that the free energy cost for CsCl crystal growth in this system is even larger.  However, similar
 to the melting study the crystal that grows around the initial CsCl cluster in the $x_{B}=0.2$ system is made up of only ``A'' particles. Next we study the free energy cost for fcc crystallization in $x_{B}=0.2$ system. We find that the free energy cost to grow a fcc crystal from 15-35 cluster size in $x_{B}=0.2$ system
 is lower than the free energy cost to grow a similar size range CsCl crystal for $x_{B}=0.3$. This implies that in the $x_{B}=0.2$ system the free energy cost for fcc crystallization is lower than the CsCl crystallization. Note that although we make this comparative statement we are unable to determine the free energy cost for growing a CsCl crystal in the $x_{B}=0.2$ which leads us to believe that the cost must be very high.

\begin{figure}[h]
\centering
\includegraphics[width=.45\textwidth]{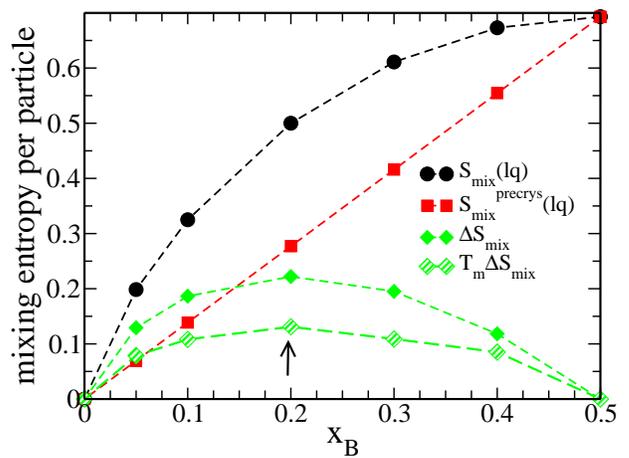}
\caption{ The mixing entropy of the liquid, $S_{mix}(lq)$, that of the partially demixed pre-crystalline liquid $S_{mix}^{precrys}(lq)$ and the difference between 
them $\Delta S_{mix}$ plotted at different compositions. We also plot the $T_{m}\Delta S_{mix}$ where $T_{m}$ is the melting temperature of the mixed CsCl+fcc crystal. 
The $\Delta S_{mix}$ shows a non monotonic composition dependence with a maxima around $x_{B}=0.2$.  } 
\label{fig7}
\end{figure}

We next analyze the origin behind the difference in the free energy cost to grow a CsCl crystal in different systems. 
In a recent study of crystallization in $Pd-Ag$ mixture it is found that the barrier to crystallization for the mixed system is about $10K_{B}T$ higher than the pure 
system \cite{jacs_demix}. The $Pd$ and the $Ag$ have a small difference in their sizes and form fcc crystal structure. Thus unlike structural frustration between the CsCl
 and fcc crystal present in the systems studied here \cite{harowell,atreyee} there exists no structural frustration in the $Pd-Ag$ system. However due to higher $Pd-Pd$ 
interaction the crystal nucleus for the $Pd-Ag$ system has a higher concentration of the $Pd$ molecules compared to that in the bulk. This leads to demixing in the system and the authors 
concluded that this demixing leads to higher barrier. In a separate study it is shown that the phase that nucleates easily is the one which has composition closer
 to the liquid \cite{monson}. In this present study we note that in the CsCl crystallization process, except for the equimolar mixture the composition of the nucleus is different from that of 
the liquid. The difference increases as we go towards smaller $x_{B}$ values. Thus it is obvious that the growth of CsCl crystal leads to demixing in the system. However, we would like to investigate if signature of demixing is present in the liquid which surrounds the crystal. 

In a recent study it has been shown that the liquid in the crystal/liquid interface shows some compositional ordering \cite{harowell_cuzr_nature_let}. In a similar 
spirit we now look at the crystal/liquid interface and investigate if the demixing takes place in the pre-crystalline liquid. For this study we perform $NP_{Z}T$ 
calculation in a rectangular box where $P_{Z}=0.5$. Initially the system consists of 432  CsCl crystal particles (equal amount of ``A" and ``B" particles) and 864 
liquid particles. The 010 layer of the crystal faces the liquid where the last layer of the crystal on one side has ``A" particles and on the other side has ``B" 
particles.  The box length in the $x$ and $y$ direction is $6.92 \sigma_{11}$. The box length in the $z$ direction is $20.76 \sigma_{11}$. Period
 boundary condition is applied in all directions. Since we want to study the interface property it is important to not have a rugged interface thus the study is 
performed above the melting temperature of the pure CsCl crystal in the respective liquid ($1.2T_{m}$) by pinning the crystal particles. Although performed above the melting temperature while equilibrating the  $x_{B}=0.38$ system we find the growth of a layer of particle on both sides of the crystal.
  In the analysis we consider these two layers, which are not pinned, to be part of the crystal. Thus for this system after equilibration there are 504 crystal particles
 and 792 liquid particles. The liquid particles span over more than $13\sigma_{11}$ distance which makes it possible to study both the interfacial and bulk properties 
of the liquid. For the $x_{B}=0.3$ system an extra layer of ``A" particles grow on the surface which has ``B" particles facing the liquid. Due to the scarcity of ``B" 
particles no extra ``B" layer grows on the other side.
In this analysis we consider the surface where the extra layer of ``A" particle has grown. We calculate the fraction of ``B" particles, $f_{B}$, within each slab of width $1 \sigma_{11}$, as a function of distance from the interface.
 The first point (z=0) in this plot is taken within the crystal which for the both the systems show same value of  $f_{B}$. Interestingly we find that for the $x_{B}=0.38$ system the concentration of the ``B" particles are higher at the interface and it gradually reaches the
 bulk value around z=4. However for $x_{B}=0.3$ system the concentration of the ``B" particles are same at the interface and at the bulk.  Thus we show that the liquid which has a lower free energy cost for crystal growth also undergoes a pre-crystalline demixing in the liquid phase. Similar to the earlier study \cite{harowell_cuzr_nature_let} we find that the
 liquid/crystal interface properties differ for apparently similar systems with different glass forming ability.

Thus we show that the process of crystallization requires demixing which takes place in the pre-crystalline liquid. We now analyze the role of demixing 
in the free energy barrier. Note that the per particle mixing entropy in a liquid can be written as,

\begin{equation}
S_{mix}(lq)=-\sum x_{i}\ln{x_{i}}
\end{equation}
\noindent
where $x_{i}$ is the mole fraction of the components.
To form CsCl+fcc crystal the liquid needs to demix. We show here that the demixing takes place in a liquid state (refer to Fig. \ref{fig6}). Although the demixing 
process happens step wise here we calculate the total effect of demixing. Thus we consider that to  form a CsCl+fcc crystal, part of the liquid needs to form a
 equimolar mixture and the other part should have pure ``A" particles. Thus the per particle mixing entropy in the pre-crystalline partially demixed liquid should be, 
\begin{equation}
S_{mix}^{precrys}(lq)=-2.0x_{B} \sum x_{i}\ln{x_{i}}
\end{equation}
\noindent     

The difference between these two entropies, $\Delta S_{mix}=S_{mix}(lq)-S_{mix}^{precrys}(lq)$, is the mixing entropy at per particle level that a liquid will   
loose in the process of partial demixing. $\Delta S_{mix}$ as a function of $x_{B}$ is shown in Fig. \ref{fig5} which shows a non-monotonic behaviour with a 
maximum around $x_{B}\simeq0.2$. Note that this kind of non monotonic behaviour is obtained in the free energy barrier to crystallization for the $Pd-Ag$ mixture which 
as discussed earlier is attributed to the demixing process\cite{jacs_demix}. 
Thus our demixing entropy study can explain the increase in the free energy cost for CsCl crystal growth with decrease in $x_{B}$ till it reaches a value of $0.2$. However this study does not
 explain why in the $x_{B}=0.2$ system where CsCl+fcc is the global minima the free energy cost for fcc crystllization is much lower  than cost for the CsCl crystallization, the latter being so high that an estimation of it is beyond the scope of the present study.

\subsection{Analysis from Energetics}
\begin{figure} [h]
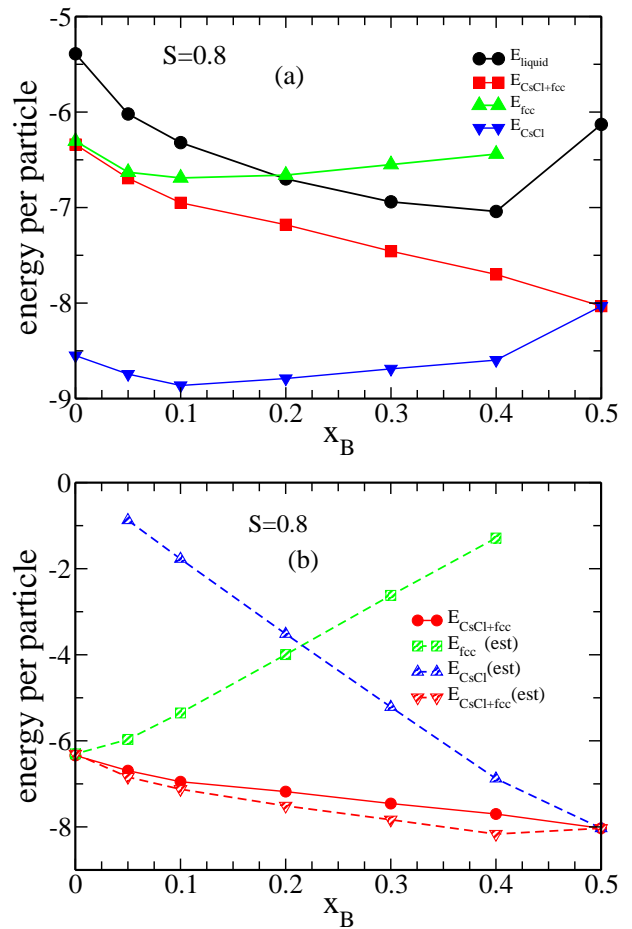

\centering
\subfigure{
\includegraphics[width=0.45\textwidth]{fig8a.eps}}
\subfigure{
\includegraphics[width=0.45\textwidth]{fig8b.eps}}
\caption{ (a) The energy per particle of the liquid, 
 mixed crystal (CsCl+fcc) , CsCl crystal and fcc crystal for different compositions. The calculations are done at $0.8T_{m}$ 
of each composition, where $T_{m}$ is the melting temperature obtained by heating the different mixed crystals.  (b) The energy of the mixed crystal,
 the estimated energy of the mixed crystal, $E_{CsCl+fcc}(est)$, the estimated contribution from the CsCl part, $E_{CsCl}(est)$, and that from the fcc part,
 $E_{fcc}(est)$, as a function of composition. The calculations are done at same temperature as in (a).}
\label{fig8}
\end{figure}

In order to understand the origin behind lower free energy cost for fcc crystallization we analyze the role of different crystal structures in crystallization by studying the energetics.
  In Fig.\ref{fig8}a we plot the energy per particle of the liquid, the mixed crystal, the fcc crystal and the CsCl crystal for different compositions at 0.8 times the
 melting temperature of their respective mixed crystals (given in Fig.\ref{fig1}). This is the melting temperature which has been obtained by step wise heating the mixed 
crystal.
  We find that the energy of the mixed crystal is always lower than the supercooled liquid, which implies that the liquid is in a metastable state. The energy of the CsCl
 crystal is always lower than the liquid. However for higher $x_{B}$ values the energy 
of the fcc crystal is above the liquid and at lower $x_{B}$ values although it becomes less than the liquid it is always higher than the CsCl value. This would imply that
 the CsCl crystal always drives the crystallization process. However this does not explain why both in the melting study and the free energy barrier calculation at
 $x_{B}=0.2$ although we can not grow CsCl crystal we can grow fcc crystals. Our subsequent analysis will explain this discrepancy.

Next we make an estimation of energy of the mixed crystal, $E_{CsCl+fcc}(est)$, at the per particle level, at different compositions by assuming that 2$x_{B}$ of
 the crystal forms CsCl and the rest forms fcc. 

\begin{eqnarray}
E_{CsCl+fcc}(est)&=&E_{CsCl}(est)+E_{fcc}(est) \nonumber\\
&=&2x_{B}E_{CsCl}+(x_{A}-x_{B})E_{fcc}
\end{eqnarray}
\noindent
Here $E_{CsCl}$ and $E_{fcc}$ are the energy of the CsCl and fcc crystal respectively, at per particle level calculated for each system at their respective 
$0.8T_{m}$. $E_{CsCl}(est)$ and $E_{fcc}(est)$ are the estimated contribution from the respective CsCl and fcc crystal part of the mixed crystal again presented at
 the per particle level.  Note that the values of $E_{CsCl}(est)$ and $E_{fcc}(est)$ take into account the fraction of the system which is in different crystal form. 
In this calculation we of course make some mistake by neglecting the surface energy. However we find that the value of energy per particle of the mixed crystal thus 
calculated is not too different from the value of the actual crystal (Fig.\ref{fig8}b). These are again calculated at the same temperatures as reported in Fig.\ref{fig8}a. 
We now break up the contribution of the two components, the contribution from CsCl and that from fcc and plot them separately. Once we do that we find that
 although at higher $x_{B}$ values the CsCl formation drives the crystallization at lower $x_{B}$ values it is the fcc crystallization which drives the crystallization.
 Although the energy per particle of the fcc crystal is still lower than that of the CsCl, the larger fraction of the fcc crystal wins over. A cross over happens just
 above $x_{B}=0.2$. This explains why the system at $x_{B}=0.2$, whose global minima is the mixed crystal, shows higher tendency towards fcc formation.

\begin{figure}[h]
\centering
\includegraphics[width=.45\textwidth]{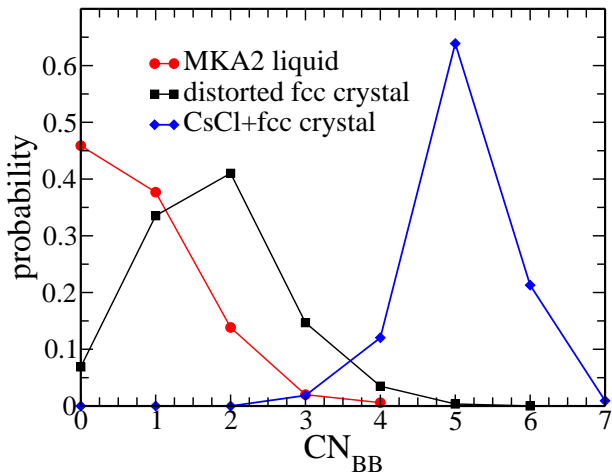}
\caption{ The probability distribution of the $CN_{BB}$ as obtained in the MKA2 liquid at $T=0.4$, the distorted fcc crystal formed by the MKA2 liquid at $T=0.35$ and CsCl+fcc crystal at $T=0.4$. The CsCl+fcc is formed at the same composition as the MKA2 liquid. The demixing required by the CsCl+fcc liquid is much higher than the distorted fcc.}
\label{fig9}
\end{figure}

However, this does not explain why the the crystallization process when driven by fcc formation has a lower free energy cost than when driven by CsCl formation. In order to understand this, we study the coordination number between the ``B" particles, $CN_{BB}$ in the $x_{B}=0.2$ system, before and after crystallization. Since we cannot crystallize the $x_{B}=0.2$ system we study the crystallization of the MKA2 model (referred earlier in the melting temperature study)  which according to Dyre and coworkers is similar in structure as the KA model but with a lower viscosity \cite{dyre}.  Confirming their conclusion we find that the LFS of the MKA2 model appears quite similar to the KA model however the dynamics is orders of magnitude faster. We now analyze the $CN_{BB}$ as obtained in the MKA2 system when it is in liquid form at $T=0.4$ and when it forms distorted fcc crystal around $T=0.35$. These are plotted in Fig. \ref{fig9}. For comparison we also plot the $CN_{BB}$ for the pure CsCl+fcc crystal at $T=0.4$. Note that the probability distribution of $CN_{BB}$ in the CsCl+fcc crystal should ideally have a peak at 6 but the peak is shifted to smaller value due to the presence of a large number of surface layer of ``B" particles. The study shows that to form distorted fcc crystal although there is an increase in the $CN_{BB}$ it is not as much as required for the CsCl+fcc crystal. Thus demixing in the distorted fcc is much weaker that CsCl+fcc. Analysis of the same kind for the $x_{B}=0.9$ system (not shown here)  shows similar behaviour.

Thus on the right hand side of the crossover where CsCl drives the crystallization there should be free energy barriers due to
 demixing. However on the left hand side the system can avoid or reduce the loss of mixing entropy by paying some energetic penalty to form distorted crystals.
 The lower energetic stability of the distorted fcc is evident from Fig.\ref{fig1}. We find that for $x_{B}=0.2$ the disordered fcc structure melts at a lower temperature compared to the mixed CsCl+fcc structure.  

\begin{figure} [h]
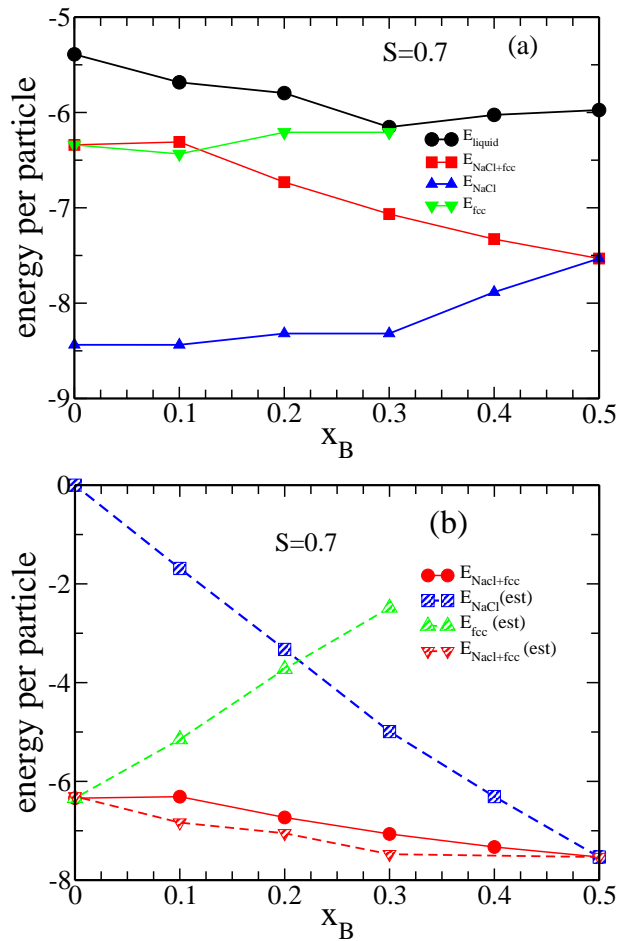

\centering
\subfigure{
\includegraphics[width=0.45\textwidth]{fig10a.eps}}
\subfigure{
\includegraphics[width=0.45\textwidth]{fig10b.eps}}
\caption{ (a) The energy per particle of the liquid,  mixed crystal (NaCl+fcc) , NaCl crystal and fcc crystal for different compositions.
 The calculations are done at $0.8T_{m}$ of each composition, where $T_{m}$ is the melting temperature obtained by heating the different mixed crystals. 
 (b) The energy of the mixed crystal, the estimated energy of the mixed crystal, $E_{NaCl+fcc}(est)$, the estimated contribution from the NaCl part, $E_{NaCl}(est)$, 
and that from the fcc part, $E_{fcc}(est)$, as a function of composition. The calculations are done at same temperature as in (a).}
\label{fig10}
\end{figure}

In order to strengthen our argument that it is indeed the demixing that frustrates the CsCl  driven crystallization process and leads to high free energy cost, we present a study of a similar system. 
Reported in a earlier study by some of us we have shown that for the NaCl system ($\sigma_{12}=0.7$) the crystallization takes place not only at equimolar 
composition but also at smaller value of $x_{B}$ forming mixed NaCl+fcc crystal \cite{atreyee}. A similar energetic study of the $\sigma_{12}=0.7$ system is shown in Fig.\ref{fig10}a.  We find that energy of the NaCl crystal is always lower than the fcc crystal (Fig.\ref{fig9}a)). A similar crossover is also obtained for this system  where
 at higher $x_{B}$ values the crystallization is driven by NaCl and at lower $x_{B}$ values it is driven by fcc (Fig. \ref{fig10}b).  Thus we should expect a similar 
crystallization problem in this system which appears not to be the case.

\begin{figure} 
\centering
\includegraphics[width=.32\textwidth]{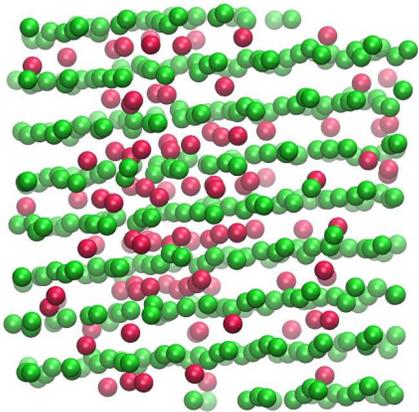}
\caption{ Snapshot of a NaCl+fcc crystal structure at $x_{B}=0.2$. The simulation is done at T=0.6 . The snapshot shows seamless formation 
of Nacl and fcc structure with no demixing. }
\label{fig11}
\end{figure}

Although the NaCl and CsCl systems appear quite similar there are some basic differences.  The CsCl crystal is made up of two interpenetrating 
sc structures of ``A" and ``B" type of particles. Thus in the CsCl+fcc crystal the ``A" particles have two different population one which forms sc and the other 
which forms fcc structure. In an earlier work we had mentioned that this wide difference in the order parameter of the two population causes the frustration between the 
two structures\cite{atreyee}. If we are away from the equimolar mixture the growth of a CsCl will deplete the population of the ``B" particles in the neighbourhood 
which should promote the formation of fcc structure between the ``A" particles. However a unit cell of fcc is not compatible with the CsCl structure thus to reduce the 
structural frustration the system sacrifices the mixing entropy and increase the concentration of the ``B" particles in the liquid near the cluster as seen in Fig.\ref{fig6} to form more CsCl structures till finally it is devoid of any more ``B'' particles in the liquid. This is the reason we find ``AB" and ``A" rich zone separated
 in Fig.\ref{fig4}a. 
The NaCl crystal on the other hand is compatible with a fcc crystal as both require the ``A" particles to form fcc structure with same lattice spacing. 
Thus unlike CsCl and fcc the NaCl and fcc can grow in a seamless fashion and the system does not require any demixing which reduces the free energy barrier. 
A snapshot of the NaCl+fcc structure is shown in Fig.\ref{fig11} which shows that there is no specific ``A" rich zone.

The study of the energetics can also explain the glass forming ability of some systems which has been previously proposed by Dyre and coworkers \cite{dyre}.
 In the above calculation if we decrease the interaction between the ``AB" particles then the contribution from the CsCl in lowering the system energy will
 decrease and the crossover will happen at a higher $x_{B}$ value. Thus the $x_{B}=0.2$ will show higher tendency of fcc crystallization as has been reported earlier
 \cite{dyre}. In the same system if we make the interaction between the ``A" particles repulsive then in a similar fashion the crossover will shift to lower $x_{B}$
 values and this will imply that the $x_{B}=0.2$ system will still be driven by the CsCl crystallization. Since this will also require demixing thus the system will
 be a better glass former as reported earlier \cite{dyre}.

\subsection{Glass Forming Ability- Role of Demixing and Eutectic Point}

We find that the loss of mixing entropy is maximum for $x_{B}=0.2$. In the free energy study within the scope of our calculation we can not grow a CsCl crystal and thus can not estimate the free energy cost to grow a CsCl crystal in this system. Which implies that free energy cost is high and w.r.t CsCl formation the $x_{B}=0.2$ system is most frustrated and a better glass former. However in the free energy calculation and the study of energetics this system shows a tendency towards fcc crystallization. The fcc crystallization also has a free energy cost because although without clear demixing the system can crystallize in fcc structure the presence of the LFS  centered around ``B" particles can frustrate this crystallization process.   
However the cost of free energy to form a  fcc crystal in $x_{B}=0.2$ system is lower than the cost of free energy to form a CsCl crystal in $x_{B}=0.3$ system. Thus it is tempting to comment that the $x_{B}=0.3$ system is a better glass former. 
However the process of crystallization is not only dependent on the free energy barrier but also on the dynamics of the system. This is the reason the eutectic point is expected to be a better glass forming region and our study of dynamics shows that indeed the $x_{B}=0.2$ system is the slowest. For the study of the dynamics we calculate $\tau_{\alpha}$ from overlap function at the respective 0.8$T_{m}$. Note these are the temperatures where the free energy calculations are done. We find that for the $x_{B}=0.38$ system $\tau_{\alpha}=17.5$  at $T=0.52$, for the $x_{B}=0.3$ system, $\tau_{\alpha}= 1350 $ at $T=0.42$ and for $x_{B}=0.2$ system $\tau_{\alpha}=4954 $ at $T=0.36$.  Thus according to the study of the dynamics the $x_{B}=0.2$ system is a better glass former. Note that the MKA2 model which undergoes crystallization differs from the KA model not in terms of the structure of the liquid but in terms of dynamics. The local structure around the smaller ``B" particles which actually frustrates the fcc crystallization is present even in the MKA2 model. However the relaxation timescale of the MKA2 model is orders of magnitude faster than the KA model. Thus our study confirms that as stated earlier\cite{dyre} it is indeed the dynamics/viscosity of the system which makes KA model a good glass former.


 \section{Conclusion}

In this article we study the comparative glass forming ability of different binary systems. In an earlier study by some of us we have shown that binary systems which form CsCl crystals in a equimolar mixture fails to crystallization if the mole fraction of the larger particles are increased \cite{atreyee}. The well known KA model is one of the systems. Thus the KA models stability against crystallization is more generic and is similar to systems which form equimolar CsCl crystal. The global structure for these systems are a mixed form of CsCl+fcc crystal \cite{harrowell-jcp}. In the CsCl+fcc crystal the bigger ``A" particles need to create two different population one which contributes towards the CsCl formation and the other which contributes towards the fcc formation. The order parameters such a BOO and coordination number of the ``A" particles are quite different in these two crystal form. Thus the failure to crystallize has been attributed to the frustration between the CsCl and fcc crystal structure.  Note that there is an array of systems which have similar frustration. However the glass forming ability of these systems although have not been calculated but is believed to be different. Thus there should be more factors contributing to the glass forming ability. 

In this article we perform a comparative study of binary glass forming liquids all having good glass forming ability and similar global minima. The study has been performed by changing the composition. We find that the free energy cost to grow a CsCl nucleus increases as we move away from an equimolar mixture. 
The study of the liquid at the liquid/crystal interface shows that the system which has lowest free energy cost to form a nucleus  also shows a demixing near the crystal surface. We believe that the structural frustration between the CsCl and fcc structure makes this demixing a prerequisite for crystallization. Our calculation of the partial demixing entropy in the liquid state shows a non monotonic dependence on composition. It shows a maxima for $x_{B}=0.2$ system. We could show a connection between the change in free energy cost to create a crystal nucleus and the change in demixing entropy as a function of composition. Our study shows that although the $x_{B}=0.2$ system is strongly frustrated against CsCl crystallization, it has tendency towards fcc growth. We can justify this tendency of fcc growth from the study of the energetics. We show that in the composition range studied here there are two regions, one which is driven by the CsCl crystallization and the other at lower $x_{B}$ values is driven by fcc crystallization.  It is primarily in the former region 
that the structural frustration between the CsCl and fcc structure leads to the requirement of demixing which eventually increases the free energy barrier and provides stability against crystallization. This point has been confirmed by studying a NaCl+fcc system which naturally undergoes crystallization at all compositions. The study of the energetics of this system also shows two similar region. However unlike the CsCl+fcc system, in the region where crystallization is driven by NaCl, due to the compatibility of the NaCl and fcc structure no demixing has been observed and the crystal grows in a seamless fashion.  
In the second region driven by fcc crystallization we show that demixing is not a stringent criteria and the stability against crystallization comes from the frustration caused by the presence of the ``B" particles with well defined LFS and also the systems proximity to eutectic point where the dynamics is slow.  Thus although we study three very similar glass former, which ideally belong to the same class of system and differ only in composition, we find that they do not share the same origin of stability against crystallization. 

We should also comment that our search of crystal structures is not exhaustive and the system which we claim to be a better glass former can crystallize in a different crystal form like the $Al_{2}Cu$ structure is found to be a low energy state of a system belonging to the same class where $x_{B}=0.33$\cite{paddyal2cu}.  This system also is known to show resistance towards crystallization. The CuZr liquid which has a low energy CsCl like structure is also a good glass former \cite{harowell_cuzr_nature_let}. Note that in these two systems the composition of the crystal is identical  to that of the liquid. Thus even above $x_{B}>0.2$ it is not always demixing which provides stability against crystallization. 








\begin{thebibliography}{53}
\makeatletter
\providecommand \@ifxundefined [1]{%
 \@ifx{#1\undefined}
}%
\providecommand \@ifnum [1]{%
 \ifnum #1\expandafter \@firstoftwo
 \else \expandafter \@secondoftwo
 \fi
}%
\providecommand \@ifx [1]{%
 \ifx #1\expandafter \@firstoftwo
 \else \expandafter \@secondoftwo
 \fi
}%
\providecommand \natexlab [1]{#1}%
\providecommand \enquote  [1]{``#1''}%
\providecommand \bibnamefont  [1]{#1}%
\providecommand \bibfnamefont [1]{#1}%
\providecommand \citenamefont [1]{#1}%
\providecommand \href@noop [0]{\@secondoftwo}%
\providecommand \href [0]{\begingroup \@sanitize@url \@href}%
\providecommand \@href[1]{\@@startlink{#1}\@@href}%
\providecommand \@@href[1]{\endgroup#1\@@endlink}%
\providecommand \@sanitize@url [0]{\catcode `\\12\catcode `\$12\catcode
  `\&12\catcode `\#12\catcode `\^12\catcode `\_12\catcode `\%12\relax}%
\providecommand \@@startlink[1]{}%
\providecommand \@@endlink[0]{}%
\providecommand \url  [0]{\begingroup\@sanitize@url \@url }%
\providecommand \@url [1]{\endgroup\@href {#1}{\urlprefix }}%
\providecommand \urlprefix  [0]{URL }%
\providecommand \Eprint [0]{\href }%
\providecommand \doibase [0]{http://dx.doi.org/}%
\providecommand \selectlanguage [0]{\@gobble}%
\providecommand \bibinfo  [0]{\@secondoftwo}%
\providecommand \bibfield  [0]{\@secondoftwo}%
\providecommand \translation [1]{[#1]}%
\providecommand \BibitemOpen [0]{}%
\providecommand \bibitemStop [0]{}%
\providecommand \bibitemNoStop [0]{.\EOS\space}%
\providecommand \EOS [0]{\spacefactor3000\relax}%
\providecommand \BibitemShut  [1]{\csname bibitem#1\endcsname}%
\let\auto@bib@innerbib\@empty



\bibitem{kob_binder}
{\sc K.~Vollmayr}, {\sc W.~Kob}, and {\sc K.~Binder},
\newblock {\em J. Chem. Phys.} {\bf 105}, 4714 (1996).

\bibitem{Acta_mater}
{\sc A.~Inoue},
\newblock {\em Acta Mater.} {\bf 48}, 279 (2000).

\bibitem{Hoover_Ree}
{\sc W.~G. Hoover} and {\sc F.~H. Ree},
\newblock {\em J. Chem. Phys} {\bf 49} (1968).

\bibitem{kob}
{\sc W.~Kob} and {\sc H.~C. Andersen},
\newblock {\em Phys. Rev. E} {\bf 51}, 4626 (1995).

\bibitem{wahnstrom}
{\sc G.~Wahnstr\"om},
\newblock {\em Phys. Rev. A} {\bf 44}, 3752 (1991).

\bibitem{NTW}
{\sc D.~Coslovich} and {\sc G.~Pastore},
\newblock {\em J. Phys.: Cond. Mat.} {\bf 21}, 285107 (2009).

\bibitem{ohern}
{\sc K.~Zhang}, {\sc Y.~Liu}, {\sc J.~Schroers}, {\sc M.~D. Shattuck}, and {\sc
  C.~S. O’Hern},
\newblock {\em 044503} {\bf 142}, 104504 (2015).

\bibitem{harowell_cuzr_nature_let}
{\sc C.~Tang} and {\sc P.~Harrowell},
\newblock {\em Nat Mater} {\bf 12}, 507 (2013).

\bibitem{harrowell-jcp}
{\sc J.~R. Fernandez} and {\sc P.~Harrowell},
\newblock {\em J. Chem. Phys.} {\bf 120}, 9222 (2004).

\bibitem{paddyal2cu}
{\sc P.~Crowther}, {\sc F.~Turci}, and {\sc C.~P. Royall},
\newblock {\em J. Chem. Phys} {\bf 143}, 044503 (2015).

\bibitem{dyre}
{\sc S.~Toxvaerd}, {\sc U.~R. Pedersen}, {\sc T.~B. Schroder}, and {\sc J.~C.
  Dyre},
\newblock {\em J. Chem. Phys.} {\bf 130}, 224501 (2009).

\bibitem{pedersen}
{\sc J.~C.~D. U.~R.~Pederson, N. P.~Bailey} and {\sc T.~B. Schroder},
\newblock {\em arXiv:0706.0813v2} .

\bibitem{frank}
{\sc F.~C. Frank},
\newblock {\em Proc. R. Soc. A} {\bf 215}, 43 (1952).

\bibitem{tarjus}
{\sc D.~Kivelson}, {\sc S.~A. Kivelson}, {\sc X.~Zhao}, {\sc Z.~Nussinov}, and
  {\sc G.~Tarjus},
\newblock {\em Physica A} {\bf 219}, 27  (1995).

\bibitem{tarjus_PRL_curvedspace}
{\sc F.~m.~c. Sausset}, {\sc G.~Tarjus}, and {\sc P.~Viot},
\newblock {\em Phys. Rev. Lett.} {\bf 101}, 155701 (2008).

\bibitem{tanaka-jnoncryst2}
{\sc H.~Tanaka},
\newblock {\em J. Non-Crys. Solids} {\bf 351}, 3371  (2005).

\bibitem{tanaka-nature12}
{\sc M.~Leocmach} and {\sc H.~Tanaka},
\newblock {\em Nat Commun} {\bf 3}, 974 (2012).

\bibitem{tanaka-jnoncryst1}
{\sc H.~Tanaka},
\newblock {\em J. Non-Crys. Solids} {\bf 351}, 3385  (2005).

\bibitem{vlot}
{\sc M.~J. Vlot}, {\sc H.~E.~A. Huitema}, {\sc A.~de~Vooys}, and {\sc J.~P.
  van~der Eerden},
\newblock {\em J. Chem. Phys.} {\bf 107}, 4345 (1997).

\bibitem{tanaka-epje}
{\sc H.~Tanaka},
\newblock {\em Eur. Phys. J. E} {\bf 35}, 113 (2012).

\bibitem{tanaka-vshaped-prl}
{\sc M.~Kobayashi} and {\sc H.~Tanaka},
\newblock {\em Phys. Rev. Lett.} {\bf 106}, 125703 (2011).

\bibitem{valeria}
{\sc V.~Molinero}, {\sc S.~Sastry}, and {\sc C.~A. Angell},
\newblock {\em Phys. Rev. Lett.} {\bf 97}, 075701 (2006).

\bibitem{charu_vshaped}
{\sc D.~Nayar} and {\sc C.~Chakravarty},
\newblock {\em Phys. Chem. Chem. Phys.} {\bf 15}, 14162 (2013).

\bibitem{atreyee}
{\sc A.~Banerjee}, {\sc S.~Chakrabarty}, and {\sc S.~M. Bhattacharyya},
\newblock {\em J. Chem. Phys.} {\bf 139}, 104501 (2013).

\bibitem{harowell}
{\sc J.~R. Fern\'andez} and {\sc P.~Harrowell},
\newblock {\em Phys. Rev. E} {\bf 67}, 011403 (2003).

\bibitem{valdes-jcp}
{\sc L.~C. Valdes}, {\sc F.~Affouard}, {\sc M.~Descamps}, and {\sc
  J.~Habasaki},
\newblock {\em J. Chem. Phys.} {\bf 130}, 154505 (2009).

\bibitem{lammps}
{\sc S.~J. Plimpton},
\newblock {\em J. Comput. Phys.} {\bf \textbf{117}}, 1 (1995).

\bibitem{steinhardt}
{\sc P.~J. Steinhardt}, {\sc D.~R. Nelson}, and {\sc M.~Ronchetti},
\newblock {\em Phys. Rev. B} {\bf 28}, 784 (1983).

\bibitem{dellago}
{\sc W.~Lechner} and {\sc C.~Dellago},
\newblock {\em J. Chem. Phys.} {\bf 129}, 114707 (2008).

\bibitem{unravel}
{\sc M.~K. Nandi}, {\sc A.~Banerjee}, {\sc S.~Sengupta}, {\sc S.~Sastry}, and
  {\sc S.~M. Bhattacharyya},
\newblock  {\bf 143}, 174504 (2015).

\bibitem{wham}
{\sc S.~Kumar}, {\sc J.~M. Rosenberg}, {\sc D.~Bouzida}, {\sc R.~H. Swendsen},
  and {\sc P.~A. Kollman},
\newblock {\em J. Comp. Chem.} {\bf 13}, 1011 (1992).

\bibitem{jacs_demix}
{\sc C.~Desgranges} and {\sc J.~Delhommelle},
\newblock {\em JACS} {\bf 136}, 8145 (2014).

\bibitem{monson}
{\sc S.~Punnathanam} and {\sc P.~A. Monson},
\newblock {\em J. Chem. Phys.} {\bf 125}, 024508 (2006).

\end{thebibliography}
 
\clearpage

\end{document}